\documentclass[superscriptaddress,aps,pra,nofootinbib,notitlepage,10pt,longbibliography,twocolumn]{revtex4-1}
\pdfoutput=1
\usepackage{graphicx}
\usepackage{amsmath}
\usepackage{amssymb}
\usepackage{amsthm}
\usepackage{comment}
\usepackage{cprotect}
\usepackage{placeins}
\usepackage[caption=false]{subfig}
\usepackage[colorlinks]{hyperref}
\usepackage[all]{hypcap}
\usepackage{tikz}
\usepackage{verbatim}
\usepackage{subfig}
\usetikzlibrary{arrows}
\usepackage{units}
\usepackage{soul}
\usepackage{textcomp}

\usepackage{listings}
\usepackage{color}
\usepackage[utf8]{inputenc}
\definecolor{codegreen}{rgb}{0,0.6,0}
\definecolor{codegray}{rgb}{0.5,0.5,0.5}
\definecolor{codepurple}{rgb}{0.58,0,0.82}
\definecolor{backcolour}{rgb}{0.95,0.95,0.92}

\lstdefinestyle{mystyle}{
  backgroundcolor=\color{backcolour},   commentstyle=\color{codegreen},
  keywordstyle=\color{magenta},
  numberstyle=\tiny\color{codegray},
  stringstyle=\color{codepurple},
  basicstyle=\footnotesize,
  breakatwhitespace=false,         
  breaklines=true,                 
  captionpos=b,                    
  keepspaces=true,                 
  numbers=left,                    
  numbersep=5pt,                  
  showspaces=false,                
  showstringspaces=false,
  showtabs=false,                  
  tabsize=2
}

\newcommand{\eq}[1]{Eq.~\hyperref[eq:#1]{(\ref*{eq:#1})}}
\renewcommand{\sec}[1]{\hyperref[sec:#1]{Section~\ref*{sec:#1}}}
\DeclareRobustCommand{\app}[1]{\hyperref[app:#1]{Appendix~\ref*{app:#1}}}
\newcommand{\tab}[1]{\hyperref[tab:#1]{Table~\ref*{tab:#1}}}
\newcommand{\fig}[1]{\hyperref[fig:#1]{Figure~\ref*{fig:#1}}}
\newcommand{\figa}[2]{\hyperref[fig:#1]{Figure~\ref*{fig:#1}#2}}
\newcommand{\figx}[2]{\hyperref[fig:#1]{Figure~\ref*{fig:#1}(#2)}}
\newcommand{\thm}[1]{\hyperref[thm:#1]{Theorem~\ref*{thm:#1}}}
\newcommand{\lem}[1]{\hyperref[lem:#1]{Lemma~\ref*{lem:#1}}}
\newcommand{\cor}[1]{\hyperref[cor:#1]{Corollary~\ref*{cor:#1}}}
\newcommand{\defn}[1]{\hyperref[def:#1]{Definition~\ref*{def:#1}}}
\newcommand{\alg}[1]{\hyperref[alg:#1]{Algorithm~\ref*{alg:#1}}}

\def\bra#1{\mathinner{\langle{#1}|}}
\def\ket#1{\mathinner{|{#1}\rangle}}

\input{Qcircuit}

\begin{document}

\title{Increasing the representation accuracy of quantum simulations of chemistry\\ without extra quantum resources}

\date{\today}
\author{Tyler Takeshita}
\email{tyler.takeshita@daimler.com}
\affiliation{Mercedes-Benz Research and Development North America, Sunnyvale, CA 94085}
\author{Nicholas C.\ Rubin}
\email{nickrubin@google.com}
\affiliation{Google Inc., Venice, CA 90291}
\author{Zhang Jiang}
\affiliation{Google Inc., Venice, CA 90291}
\author{Eunseok Lee}
\affiliation{Mercedes-Benz Research and Development North America, Sunnyvale, CA 94085}
\author{Ryan Babbush}
\affiliation{Google Inc., Venice, CA 90291}
\author{Jarrod R.\ McClean}
\email{jmcclean@google.com}
\affiliation{Google Inc., Venice, CA 90291}

\begin{abstract}
Proposals for near-term experiments in quantum chemistry on quantum computers leverage the ability to target a subset of degrees of freedom containing the essential quantum behavior, sometimes called the active space.  This approximation allows one to treat more difficult problems using fewer qubits and lower gate depths than would otherwise be possible.  However, while this approximation captures many important qualitative features, it may leave the results wanting in terms of absolute accuracy (basis error) of the representation.  In traditional approaches, increasing this accuracy requires increasing the number of qubits and an appropriate increase in circuit depth as well. Here we explore two techniques requiring no additional qubits or circuit depth that are able to remove much of this approximation in favor of additional measurements.  The techniques are constructed and analyzed theoretically, and some numerical proof of concept calculations are shown.  As an example, we show how to achieve the accuracy of a $20$ qubit representation using only $4$ qubits and a modest number of additional measurements for a hydrogen molecule.  We close with an outlook on the impact such techniques may have on both near-term and fault-tolerant quantum simulations. 
\end{abstract}

\maketitle

\section{Introduction}
Quantum computers promise to make a dramatic impact on a number of fields including optimization and materials simulation.  Since the initial proposal by Feynman to simulate quantum systems via quantum systems~\cite{Feynman1982}, there has been a wealth of developments studying these applications both theoretically and experimentally.  One particular application of note is the simulation of chemical and material systems by quantum computers, as it represents a natural application of this idea in practice.  The combination of the practical potential for quantum chemistry as well as its low overhead have made it a target for the first near-term quantum computers.

The progress in quantum chemistry on quantum computers has been extremely rapid over the last few years.  Starting from the original proposal to use quantum phase estimation for chemical problems~\cite{Aspuru-Guzik2005}, the precise costs of these algorithms and methods to reduce these costs by orders of magnitude have now been developed in detail~\cite{Whitfield2010,Wecker2014,Hastings2015,Poulin2014,BabbushTrotter,McClean2014,BabbushSparse1,BabbushLowDepth,BabbushContinuum,jiang_quantum_2018,Motta2018,Campbell2018,Low2018}.  Precise gate counts are now known for hard chemical systems in implementations on the earliest fault tolerant computers under a realistic model of error correction using the surface code~\cite{Reiher2017,Babbush2018a,Berry2019,Tubman2018,Kivlichan2019}.  Methods based on quantum phase estimation, however, are believed by some to require quantum error correction, which places their experimental implementation some time beyond the current noisy intermediate-scale quantum (NISQ) devices.

As NISQ devices progress to a level of development capable of doing computations a classical computer cannot, or quantum supremacy~\cite{Boixo2016}, the question arises of if one can do a practical application such as chemistry on such a near-term quantum computer.  Prime candidates for this possibility have been variational algorithms, such as the variational quantum eigensolver (VQE)~\cite{Peruzzo2013,McClean2015} or the quantum approximation optimization algorithm~\cite{Farhi2014}.  These methods exhibit a natural adaptation to device parameters as well as intrinsic robustness to systematic errors that make them attractive candidates.  Since their inception, they have been extended to treat excited states~\cite{McClean2016b,Santagati2016,Parrish:2019} and different problem areas~\cite{Mitarai2018,farhi2018classification}, and have been demonstrated on numerous experimental architectures~\cite{Peruzzo2013,OMalley2016b,Santagati2016,Siddiqi2017b,Hempel2018,Kandala2017}.

While hardware and theoretical developments have been rapid, quantum resources are expected to remain costly for some time.  As a result, proposals for doing quantum computing for chemical problems have focused on isolating the essential strongly correlated component of a physical system for simulation on a quantum computer.  A simple division of the system into this select subsystem is often called an active space within chemistry, and more detailed treatments may incorporate it as an impurity model as well.  Despite the ability of these methods to treat qualitative phenomena that would not otherwise be accessible, this division introduces quantitative approximations that can be unacceptable when quantitative accuracy is demanded.  Previously, when one wanted to lift this approximation, the consequence was both an increased number of qubits and gate complexity, ruling out compatibility with a NISQ device.  

Here, we discuss two methods that use no additional qubits or gate complexity to lift the active space approximation in a systematic way.  One of these procedures is a novel way to leverage the quantum subspace expansions~\cite{McClean2016b}, also known to mitigate errors~\cite{Bonet2018} and provide excited states in experiments~\cite{Siddiqi2017b}, that relies only on additional measurements.  The other involves orbital relaxations, a common and well known procedure in classical electronic structure, to remove the active space approximation and reduces circuit depth for some classes of VQE circuits.  We cost out these methods, provide theoretical justification, and show how these methods work in practice with simple numerical simulations.  A framework to develop cost effective approximations is discussed that makes use of manipulation of marginals~\cite{Rubin2018}, and we conclude with an outlook on the impact for near-term quantum simulations.

\begin{figure}[t!]
\vspace{-3mm}
\includegraphics[width=6.5cm]{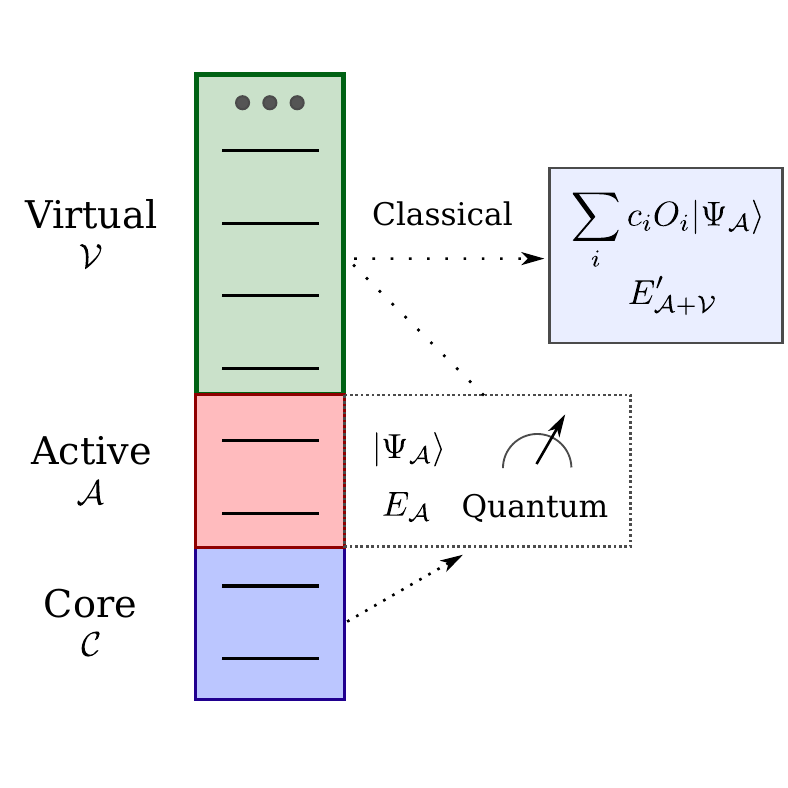}
\vspace{-3mm}
\caption{Schematic virtual quantum subspace expansion to increase accuracy without additional qubits.  The method separates the orbitals of a fermionic system into their core, active, and virtual components.  The quantum device solves the active space problem, and additional quantum measurements are taken within this space.  These additional measurements are combined with classical post-processing from data on the virtual space to correct the solutions using no additional qubits or circuit depth.  The resulting corrected wavefunction(s) are stored in a mixed quantum classical representation that may be used to derived any desired properties of these wavefunctions. \label{fig:schematic}}
\end{figure}

\section{Background and Definitions}
We begin by setting up the problem, establishing notation, and reviewing the quantum subspace expansion method~\cite{McClean2016b} as it was originally devised.  The problem of electronic structure in quantum chemistry is typically cast as the problem of determining the electronic ground state in the field of fixed nuclei, or the Born-Oppenheimer approximation~\cite{Helgaker2002}.  From here, one discretizes space, where the accuracy of this discretization determines the ultimate accuracy one can achieve.  The abstract blocks one uses to divide space are called basis functions, and a number of canonical choices are known such as linear combinations of atomic orbitals and plane waves.  Once the basis is chosen, the electronic structure Hamiltonian may be written in its canonical form as
\begin{align}\label{eq:full_space_hamiltonian}
    H = \sum_{ij} h_{ij} a_i^\dagger a_j + \frac{1}{2} \sum_{ijkl} h_{ijkl} a_i^\dagger a_j^\dagger a_k a_l
\end{align}
where each index corresponds to one of these basis functions, the $h_{ij}$ and $h_{ijkl}$ are standard integrals over the involved basis functions and the ladder operators satisfy the canonical fermionic anticommutation relations $\{a^\dagger_i, a_j\} = \delta_{ij}$, $\{a_i, a_j\} = \{a^\dagger_i, a^\dagger_j\}= 0$.  As discussed above, the number of basis functions used is tied to the ultimate accuracy one can achieve for the problem, however using too many basis functions may make the problem impractical or be a waste of resources when more clever treatments may be used.  A method that was first widely used in traditional quantum chemistry and that has been widely adopted by the quantum computing community is the active space approximation.

The physical intuition behind the active space approximation is that the space may be divided into a portion which exhibits strong correlations or entanglement, the active space, and a portion that while important, exhibits only low rank contributions that are extremely well treated perturbatively, the virtual space.  This methodology has been proven out numerous times in the classical literature by methods such as multi-reference configuration interaction and perturbation theory~\cite{Werner1988,andersson1995multiconfigurational}.  However, the size of the essential quantum component or active space, remains limited on a classical computer, and to date the contributions of virtuals have remained absent on a quantum computer.  We aim to show how one can regain virtuals on top of quantum active spaces without the need for additional qubits or gate complexity.

As depicted in Fig.~\ref{fig:schematic}, in typical chemistry calculations on quantum computers, the set of basis functions is divided into 3 sets, which we denote $\mathcal{C}$, $\mathcal{A}$, and $\mathcal{V}$ for core, active, and virtual.  The core orbitals are assumed to be doubly occupied, and their contributions are integrated out to an effective field felt by the active space and virtual space.  The virtual orbitals are ignored, and the problem is solved exactly within the dressed active space, $\mathcal{A}$.  

Our approach will utilize some of the machinery of an approach designed originally to provide excited states and error mitigation within the context of VQE, which utilizes quantum subspace expansions (QSE)~\cite{McClean2016b}.  These techniques leverage the ability to expand and manipulate representations of operators within a subspace using measurements and classical computation, without knowing the details of the states themselves.  In this framework, one assumes the ability to prepare a wavefunction within the active space that we denote as $\ket{\Psi_\mathrm{ref}}$.  We then choose a set of expansion operators $\{O_i\}$ which act on this reference in combination with another operator, such as the Hamiltonian, to form a representation of that operator in the basis given by $\{O_i \ket{\Psi_\mathrm{ref}} \}$.  As this basis may be non-orthogonal, we also measure the overlap or metric matrix within this basis in order to ensure the problem is well-defined.  The matrices may be formed through additional measurements only, and have matrix elements given by
\begin{align}
    H_{ij} = \bra{\Psi_\mathrm{ref}} O_i^\dagger H O_j \ket{\Psi_\mathrm{ref}} \label{eq:H_ij}\\
    S_{ij} = \bra{\Psi_\mathrm{ref}} O_i^\dagger O_j \ket{\Psi_\mathrm{ref}} \label{eq:S_ij}
\end{align}
with these matrices, one then uses canonical diagonalization to remove the approximately $0$ eigenvalues of $S$ and solve the generalized eigenvalue problem in the well conditioned subspace given by
\begin{align}
    HC = SCE
\end{align}
where $C$ is the matrix of eigenvectors in the basis $\{O_i \ket{\Psi_\mathrm{ref}} \}$, and $E$ is the diagonal matrix of eigenvalues.  This approach can both improve the accuracy of the ground state and provide approximations to excited states.  In the original work~\cite{McClean2016b}, it was suggested to use $O_i$ approximating fermionic excitations of the form
\begin{align}
    \{a_p^\dagger a_q | p, q \in \mathcal{A} \}
\end{align}
which when considered with the Hamiltonian, composed of only up to 2-particle operators, means that the matrix elements can be evaluated as sums over subsets of the 4-electron reduced density matrix (RDM) where the sum weights are determined by integrals in the Hamiltonian.  The 4-electron density matrix in the active space is 
\begin{align}\label{eq:4-rdm}
    {}^{(4)}D_{\mathcal{A}\; pqrs}^{tuvw} = \bra{\Psi_\mathrm{ref}}a_w^\dagger a_v^\dagger a_u^\dagger a_t^\dagger a_p a_q a_r a_s \ket{\Psi_\mathrm{ref}}.
\end{align}
Each of these elements can be evaluated on a quantum computer by repeated preparation of $\ket{\Psi_\mathrm{ref}}$ and measurement of the Pauli operators corresponding to the transformation of this matrix element by a Jordan-Wigner or equivalent transformation. To perform the procedure exactly the number of terms one must measure scales as $N_A^8$ where $N_A$ is the number of active space orbitals.

\section{Virtual Quantum Subspace Expansion}

The original formulation of the QSE remained in the active space, and did not include contributions from the virtual orbitals.  Here we show how simple classical post-processing and measurements can be used to re-introduce contributions from the virtuals in a systematic way, to approach quantitative accuracy for the chemical problem.  

Consider a reference function which is constructed only on the original active space $\mathcal{A}$.  We now introduce a set of expansion operators
\begin{align}\label{eq:excitation_op}
    \big\{a_i^\dagger a_p, a_\mu^\dagger a_q a_\nu^\dagger a_r \mid i \in \mathcal{A} \cup \mathcal{V};\, p, q, r \in \mathcal{A};\, \mu, \nu \in \mathcal{V} \big\}
\end{align}
that act, in principle, on additional qubits that would define the virtual space.  However, by definition, the reference wavefunction has no components in the virtual space,  and hence the contraction over the virtual space can be done using only efficient classical computation on the Hamiltonian in addition to the information from the active space RDM. The only information that is required is the integrals over the full space, which are classical inputs to the problem, as well as the appropriate density matrix elements.

We note that the quantum advantage over classical multi-reference configuration interaction (MRCI) methods here is that the reference wavefunction may contain an exponential number of determinants, which gives classical MRCI methods exponential scaling in the size of the active space.  In contrast, our method scales only polynomially in the size of the active space, and if one measures the appropriate RDM of the active space, the quantum computing resources required are independent of the size of the virtual space. For both classical MRCI and the method proposed here, the classical computation required scales polynomially in the size of the virtual space. Thus, our method allows for very large basis sets to be treated on a small quantum computer.

In this set of operators, we have restricted ourselves to single and double excitations from the references within the active space and virtuals.  The single and double excitations have empirically been shown to be the dominant contributions in classical multireference methods.  Our method is general enough to go to higher order approximations, however, with each higher order of excitation, a larger reduced density matrix must be measured.  If we denote single and doubles as excitation level 2, at excitation level $k$, one must measure the $(2 + k)$-RDM, which is expected to have a cost that scales as $N_A^{2(2+k)}$ in the number of terms that must be measured.  When the excitation level $k$ matches the number of electrons, the method is formally exact.

To illustrate how we eliminate the virtual space, we take the highest order matrix element as an example.  The relevant quantity to measure for the case where both expansion operators in Eq.~\eqref{eq:H_ij} are double excitations is
\begin{align}\label{eq:both_double}
M = \bra{\Psi_\mathrm{ref}} a_\xi a_s^\dagger a_\eta a_r^\dagger\: a_i^\dagger a_j^\dagger a_k a_l\: a_\mu^\dagger  a_p a_\nu^\dagger a_q \ket{\Psi_\mathrm{ref}}\,.
\end{align}
Using Wick's theorem where Greek indices denote virtual orbitals, we contract the operators on the virtual space,
\begin{align}
  M  &= \Delta_{\xi \eta, \mu \nu}\, \big\langle a_s^\dagger a_r^\dagger a_i^\dagger a_j^\dagger a_k a_l  a_p a_q \big\rangle +\sum_{x= i, j}\,\sum_{y = l, k}\nonumber\\
  &\quad (-1)^{\delta_{x j} + \delta_{y k}} \big(\delta_{\xi x}\Delta_{y\eta, \mu \nu}- \delta_{\eta x}\Delta_{y\xi, \mu \nu} \big)\big\langle a_s^\dagger a_r^\dagger a_{\bar x}^\dagger a_{\bar y} a_p a_q \big\rangle \nonumber\\
  &\quad +\Delta_{\xi \eta, i j} \Delta_{k l, \mu \nu}\,  \big\langle a_s^\dagger a_r^\dagger a_p a_q \big\rangle\,, \label{eq:contract_double}
\end{align}
where $\Delta_{\xi \eta, \mu \nu} = \delta_{\xi\nu}\delta_{\eta\mu}-\delta_{\xi\mu}\delta_{\eta\nu}$, $\bar x$ ($\bar y$) means $x$ ($y$) changing to the other value in the sum (e.g. $\bar i = j$ for x), and $\langle \cdots \rangle$ denotes expectation values  with respect to the reference state $\ket{\Psi_\mathrm{ref}}$ in the active space.  The reduction in Eq.~\eqref{eq:contract_double} shows that the expectation value $M$ can be derived by knowing the $4$-election RDM of the reference state $\ket{\Psi_\mathrm{ref}}$ in the active space.  Similar but simpler results hold by replacing one or both double excitation operators by single excitation operators. Hence the inclusion of virtual orbitals amounts to simple classical post-processing and no additional qubits after the appropriate measurements have been performed.

\begin{figure}[t!]
\includegraphics[width=8cm]{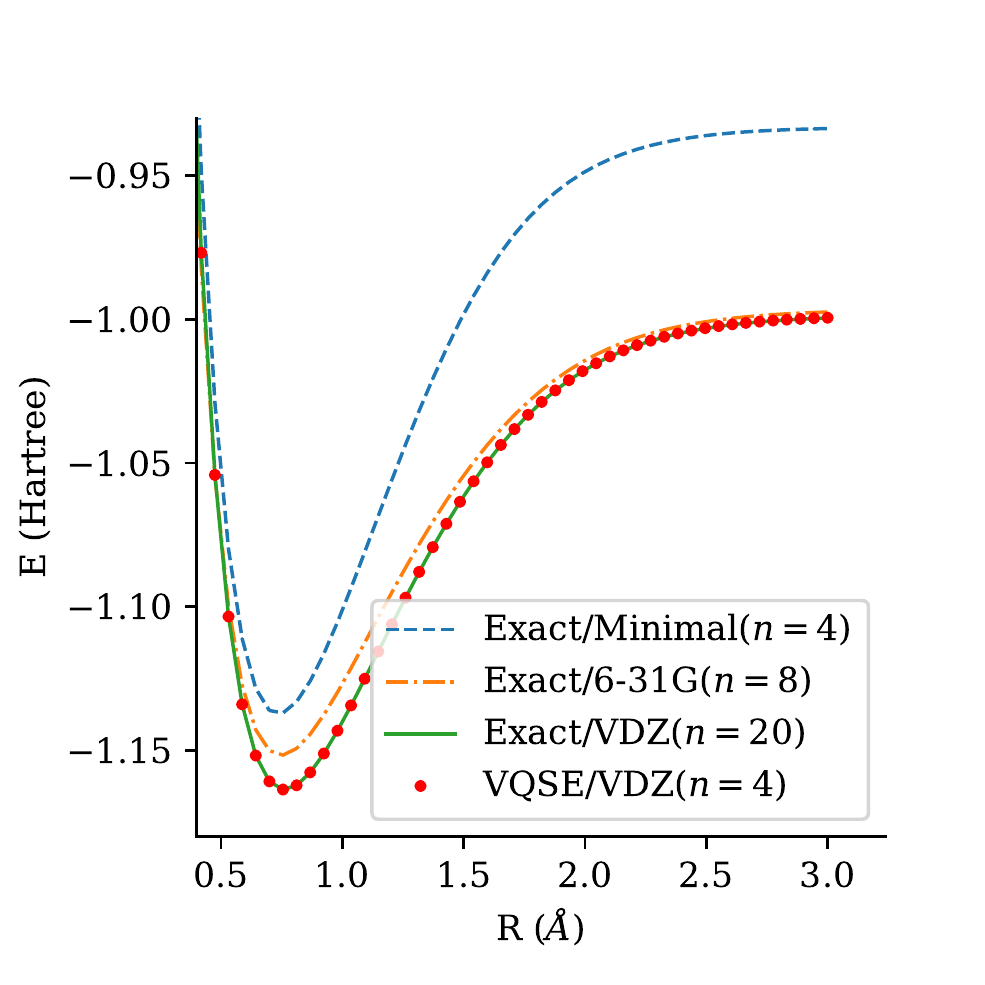}
\vspace{-3mm}
\caption{Ground state energy as a function of internuclear separation for the H$_2$ molecule at different levels of theory.  The notation $(n=n_q)$ after each line indicates that $n_q$ qubits are needed for that representation. The use of the virtual quantum subspace expansion (VQSE) technique using only 4 qubits attains the same accuracy in this case as a representation using $20$ qubits, with an error much smaller than chemical accuracy (approximately $10^{-5} E_h$). The minimal basis here denotes an STO-3G calculation and 6-31G is an intermediate level of accuracy between this and cc-pVDZ (here, VDZ).\label{fig:pes}}
\end{figure}

\section{Cumulant and Restricted Active Space Approximations}
One may introduce a number of approximations to make the construction more efficient.  The simplest approximation is the division of the active space into a part which is excited into the virtuals $\mathcal{A}_v$ and a part of the active space which will be treated as correlated core orbitals $\mathcal{A}_c$.  This reduces the scaling in the number of measurements to the cost of originally estimating the energy plus the size of the virtuals active space $\mathcal{A}_v$, $N_{Av}^8$, which for small sizes can dramatically reduce the cost.

The second class of approximations involves estimating the matrix elements of the reduced density matrix via cumulant approximations.  For example, one can form a series of approximations to the 4 RDM using products of lower RDMs and perturbative corrections.  This may be denoted by
\begin{align}
    {}^{(4)}D = & {}^{(4)}\Delta + 4 {}^{(3)}\Delta \wedge {}^{(1)}\Delta + \notag \\ 
    & 3 {}^{(2)}\Delta \wedge {}^{(2)}\Delta + 6 {}^{(2)}\Delta \wedge
    {}^{(1)}\Delta \wedge {}^{(1)}\Delta + \notag \\
    & {}^{(1)}\Delta \wedge {}^{(1)}\Delta \wedge {}^{(1)}\Delta \wedge {}^{(1)}\Delta
\end{align}
where ${}^{(l)}\Delta$ is the $l$-particle cumulant, expressing irreducible $l$-body correlations in the density matrix, and $\wedge$ is the Grassmann wedge product.  The simplest form of approximation is given by setting ${}^{(l)}\Delta = 0$ for $l>2$ which allows one to express the 4-RDM with only measurements from the original 2-RDM.  This greatly reduces the number of terms to measure back to $N_{A}^4$, but introduces some considerable approximation to the energetic values.  Much work has been done on improving these approximations as well.  An alternative scheme we do not exploit here may stochastically sample the elements to measure with hopefully increasing degrees of accuracy as time proceeds within a calculation.

\begin{figure}[t!]
\includegraphics[width=8cm]{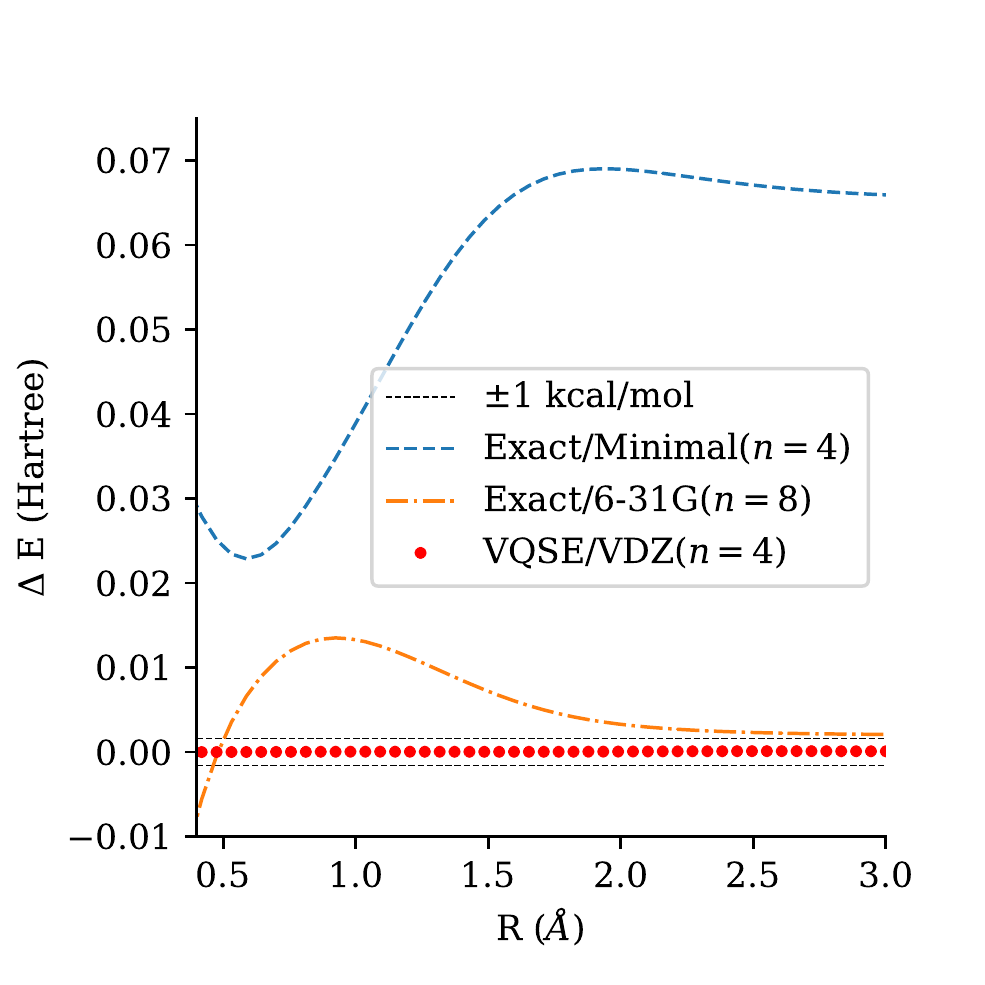}
\vspace{-3mm}
\caption{Ground state energy error with respect to the exact solution in a double zeta basis (cc-pVDZ, abbreviated here as VDZ) as a function of internuclear separation for the H$_2$ molecule for different levels of theory.  The notation $(n=n_q)$ after each line indicates that $n_q$ qubits are needed for that representation. We see that the virtual quantum subspace expansion (VQSE) technique using only 4 qubits attains the same accuracy as an exact solution in a 20 qubit basis, up to an error of $10^{-5} E_h$ which is far below chemical accuracy. \label{fig:e_diff}}
\end{figure}

\begin{figure}[t!]
    \centering
    \includegraphics[width=8cm]{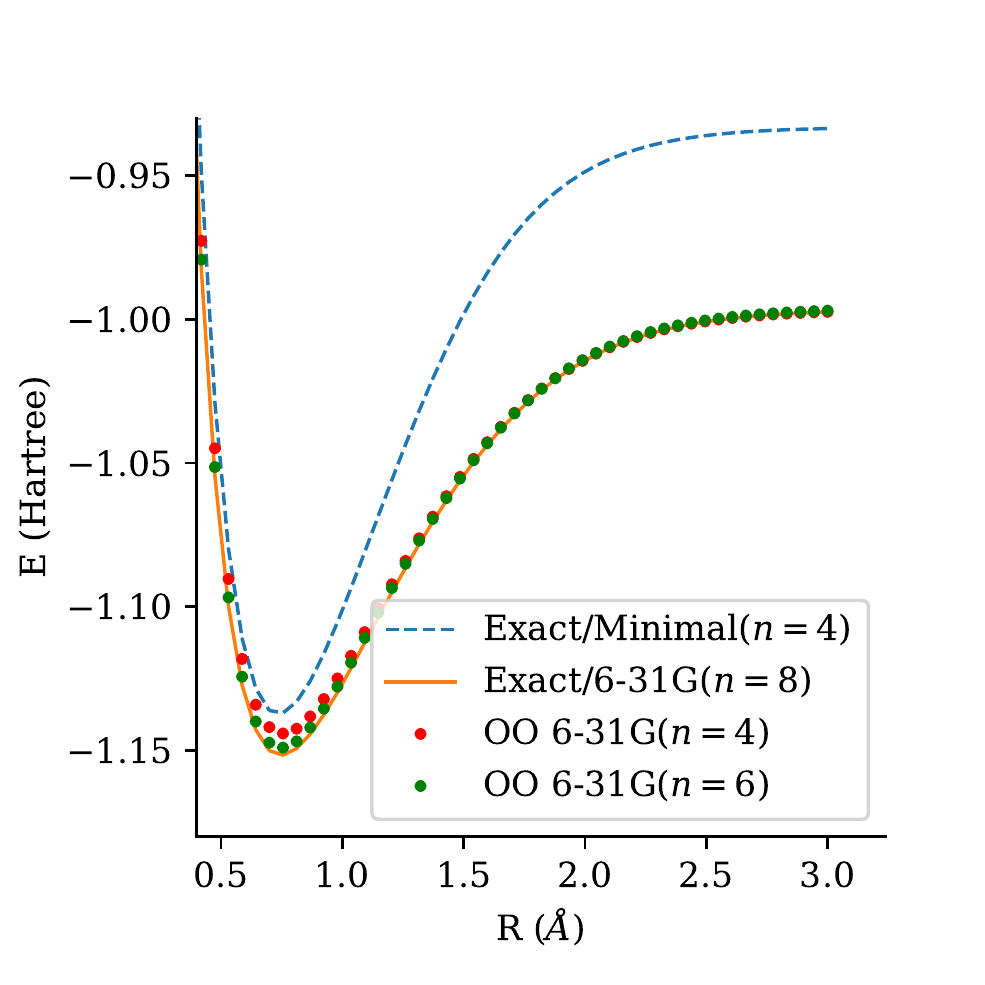}
    \vspace{-3mm}
    \caption{Ground state energy as a function of internuclear separation for the H$_2$ molecule at different levels of theory.  The notation $(n=n_q)$ after each line indicates that $n_q$ qubits are needed for that representation. The orbital optimization post-processing is labeled by ``OO".  We compare active space calculations on the full space with the 6-31G basis set (requiring 8 qubits) to active space calculations on the same model involving only 4 and 6 qubits.}
    \label{fig:oo_h2_631g}
\end{figure}
\section{Numerical Experiments}
In this section, we show on a prototype system the accuracy gains one may expect from using this method on a real system. We assess the performance of the procedure on a simple molecule, and examine the ground state energy as a function of the number of additional virtual orbitals considered in the system. The first system we look at is H$_2$ in a variety of different basis sets.  These numerics were enabled by the OpenFermion software package for doing quantum chemistry on a quantum computer~\cite{OpenFermion}.  The orbitals were obtained from a Hartree-Fock calculation, and the reference function $\ket{\Psi_\mathrm{ref}}$ was the exact solution within the active space consisting of the single occupied orbital and the lowest energy virtual orbital.  Only singles and doubles excitations from the active space to the additional virtual orbitals were included in these calculations. Correlation consistent basis set of double zeta quality (cc-pVDZ) ~\cite{Dunning1989} are used as well as the 6-31G basis set~\cite{Hehre1969} for reference.  We compare with the exact results with both basis sets.

The results of the numerical simulations are shown in Fig.~\ref{fig:pes} and the energy differences with respect to Exact/cc-pVDZ are shown in Fig.~\ref{fig:e_diff}. We see that the addition of virtuals to an active space at a higher level of theory quickly surpasses what is possible at a lower level of theory, using no additional qubits.  In fact, using only 4 qubits, which is the same as a typical minimal basis for H$_2$, the VQSE procedure attains an accuracy commensurate with the exact solution on a basis that would require $20$ qubits.  The curves smoothly improve as a function of the number of virtuals included, and show excellent accuracy across the range of calculations.

\begin{figure}[t!]
    \centering
    \includegraphics[width=8cm]{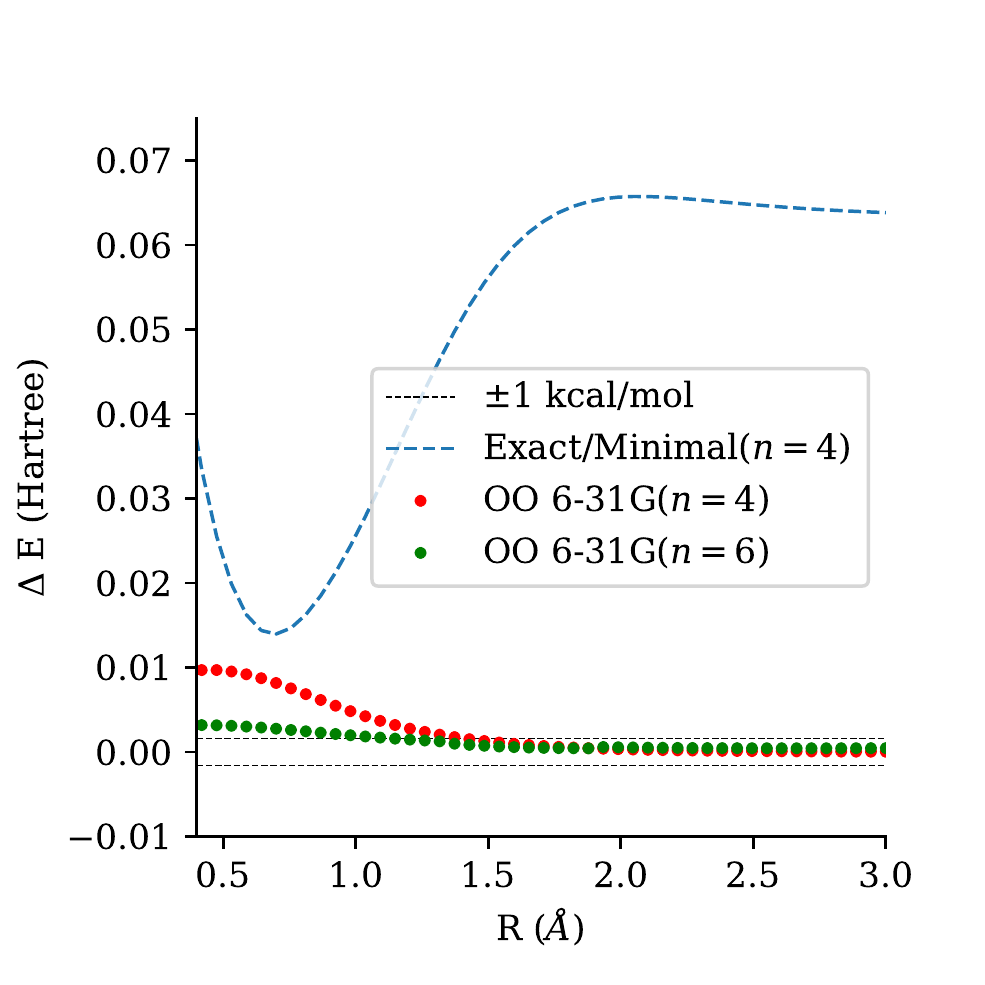}
    \vspace{-3mm}
    \caption{Ground state energy error with respect to the exact solution in a 6-31G basis as a function of internuclear separation for H$_2$.  The notation $(n=n_q)$ after each line indicates that $n_q$ qubits are needed for that representation. The orbital optimization post-processing is labeled by ``OO". We see that orbital optimization greatly reduces the need for perturbative corrections in the stretched bonding region.}
    \label{fig:my_label2}
\end{figure}

\section{Full space orbital relaxation}
A common classical electronic structure method to improve active space calculations is to allow for orbitals to relax in the presence of the newly optimized active space wavefunction. Variants of this idea fall into the family of methods known as multiconfigurational self-consistent-field (MCSCF).  These algorithms work by iteratively solving the active-space Schr\"odinger equation and then finding a single-particle rotation on the full space that minimizes the energy.   Integrating a quantum resource as an active space solver in the MCSCF framework has been previously suggested for use with the phase estimation approach to quantum simulation ~\cite{Reiher2017} and also motivated other embedding methods in the quantum computing context ~\cite{rubin2016hybrid}. Here we demonstrate how a single step of an MCSCF orbital relaxation can be used to further improve energies with a procedure that is designed for the limitations of NISQ simulations.  This technique requires measuring only the two-electron reduced-density-matrix with no additional quantum resources.

Given a $2$-RDM from the ground state of the dressed active space Hamiltonian we seek to minimize the energy of the full space Hamiltonian Eq.~\eqref{eq:full_space_hamiltonian} through one-body rotations $U$ on the full space. Due to Thouless's theorem, the one-body rotations can be efficiently implemented as a rotation of the underlying basis~\cite{thouless1960stability}.  This can be formulated as the following non-linear optimization problem:
\begin{align}\label{eq:opt_oo}
\min_{U} E = \sum_{ij}u_{i,i'}u_{j,j'}^{*}h_{ij}\langle a_{i}^{\dagger}a_{j}\rangle_{\mathcal{A}} + \nonumber \\
\sum_{ijkl}u_{i,i'}u_{j,j'}u_{k,k'}^{*}u_{l,l'}^{*}h_{ijkl}\langle a_{i}^{\dagger}a_{j}^{\dagger}a_{k}a_{l}\rangle_{\mathcal{A}} \nonumber \\
\mathrm{s.t.} \;\;Ua_{j}^{\dagger}U^{\dagger} = \sum_{j'}u_{j,j'}a_{j'}^{\dagger} \nonumber \\
U = e^{X} \nonumber \\
X = \sum_{p,q}t_{p,q}a_{p}^{\dagger}a_{q} \;\;,\;\; t_{p,q} = -t_{q, p}^* \nonumber\\
U^{\dagger}U = \mathbb{I}
\end{align}
where $\langle a_{i}^{\dagger}a_{j}^{\dagger}a_{k}a_{l}\rangle_{\mathcal{A}}$ and $\langle a_{i}^{\dagger}a_{j}\rangle_{\mathcal{A}}$ are the $2$- and $1$-RDM of the ground state active space wavefunction.

In the classical literature it is common to use the second order approximation of $u$
\begin{align}
u = e^{t} \approx \mathbb{I} + t + \frac{1}{2}t^{2}
\end{align}
to minimize energy in a Newton-Raphson style~\cite{siegbahn1981complete, siegbahn1980comparison, szalay2011multiconfiguration, fosso2016large}. 

A well known alternative to parameterizing the unitary as an exponentiated antihermitian matrix is the use of Givens rotations~\cite{head1988optimization, ivanic2003mcscf}.  This parameterization uses a set of angles $\{\theta\}$ associated with the set of non-redundant orbital rotation generators.  For $2$-RDMs obtained from an exact diagonalization of the active space Hamiltonian the only non-redundant parameters are one-body generators associated with pairs of orbitals involving rotations from the active-space to the virtual space and active-space to the core-space.  Therefore, the unitary in~\eqref{eq:opt_oo} can be expressed as a product of Givens matrices
\begin{align}
U = \prod_{i \in \mathrm{active}}\prod_{b\in \mathrm{core, virtual}}G_{i,b}(\theta_{i,b}).
\end{align}
The optimal rotation of a single angle with respect to the input $2$-RDM, one-, and two-electron integrals has been derived~\cite{ivanic2003mcscf} along with a sweep procedure to find an energy minimizing $U$. This algorithm was used for developing $2$-RDM MCSCF methods~\cite{gidofalvi2008active} and the first orbital-optimized coupled-cluster doubles methods~\cite{sherrill1998energies}.

In the context of NISQ algorithms one can iterate between solving the active space Schr\"odinger equation and full-space one-body rotations, similar to a general MCSCF routine, or perform a single orbital relaxation procedure as post-processing. Because of the variational principle, relaxing the orbitals in a single step as post-processing is guaranteed to reduced the energy.  Furthermore, if the ground state of the active space is not achieved due to an approximate wavefunction ansatz, additional one-body orbital rotations between orbital pairs inside the active space can be included in the relaxation.  Including these rotations would correspond to an additional linear depth circuit that would have been executed perfectly on the quantum computer ~\cite{PhysRevLett.120.110501}.   Any circuit that contains one-body rotations at the end can be made shorter with classical post-processing by instead rotating the operator to be measured \cite{Johnson2019APSoo}.  

To demonstrate the utility of the orbital relaxation step for NISQ devices we examine the energy improvement upon orbital relaxation for molecular hydrogen computed in the 6-31G basis set.  We select the lowest energy Hartree-Fock orbitals as the active space and perform orbital optimization with the Scipy COBYLA optimizer.  Though this is not guaranteed to find the lowest energy orbital rotations it is sufficient for validating the performance one can achieve through orbital relaxation as post-processing.  In Figure~\ref{fig:oo_h2_631g} we plot the energy of molecular hydrogen and observe a significant recovery of energy when performing a 4-qubit active space calculation with orbital relaxation on the full 8-qubit space.  The 6-qubit active space calculation is included to demonstrate a smooth interpolation of this method to exact diagonalization in the full 8-qubit space.  Figure~\ref{fig:my_label2} shows the energy difference relative to an 8-qubit exact diagonalization, revealing we reach within $5\times 10^{-5}$ $E_{h}$ of the exact solution with half the number of qubits.

\section{Discussion}

As NISQ devices progress, there will continue to be a push to demonstrate their power for interesting chemical problems.  As the number of qubits and coherence time are likely to remain limited, one expects approaches such as active space divisions of the orbitals will continue to play a dominant role.  While powerful, these techniques introduce a level of approximation that may prevent reaching the desired accuracy for problems of interest without additional qubits or gate depth.

Here we have introduced a method for going beyond the active space approximation with no additional qubits or gate complexity.  We showed how this method may be constructed and demonstrated its power for simple test systems.  While the number of measurements is increased, there are a number of promising approximation schemes that may allow one to avoid this additional overhead in an intelligent way.  Moreover, these methods maintain an exponential advantage over their classical counterparts in the treatment of the active space and active space reference.  We believe methods such as these will be crucial for obtaining accurate solutions on NISQ devices.

Additionally, in the long-term these techniques should useful for the competitiveness of fault-tolerant approaches to simulating quantum chemistry. Studies such as \cite{Reiher2017,Berry2019} have focused on performing error-corrected quantum computations of chemistry within an active space, and the methods developed here should help significantly to capture dynamic correlation in those simulations; in fact, an MCSCF procedure was even suggested in \cite{Reiher2017}. However, more research would be needed to make such schemes viable within error-correction. This is because both schemes presented here require a large (although polynomial scaling) number of measurements and error-corrected logical gates are many orders of magnitude slower than the physical gates used in NISQ devices.

\bibliographystyle{apsrev4-1_with_title}
\bibliography{Mendeley,extra_references,ryan}

\end{document}